\documentclass[onecolumn,showpacs,fleqn,nobibnotes]{revtex4}

\usepackage{amsmath}
\usepackage{graphicx}
\usepackage{float}
\usepackage{subfigure}

\def\lsim{\raise0.3ex\hbox{$<$\kern-0.75em\raise-1.1ex\hbox{$\sim$}}}
\def\gsim{\raise0.3ex\hbox{$>$\kern-0.75em\raise-1.1ex\hbox{$\sim$}}}

\def\beq{\begin{equation}}
\def\eeq{\end{equation}}
\def\beqa{\begin{eqnarray}}
\def\eeqa{\end{eqnarray}}

\newcommand{\rr}{\mbox{\boldmath $r$}}
\newcommand{\rrn}{\mbox{$r$}}

\newcommand{\rb}{\mbox{\boldmath $b$}}
\newcommand{\N}{\mathcal{N}}
\newcommand{\real}{\mathrm{Re}\,}
\newcommand{\imaginary}{\mathrm{Im}\,}
\newcommand{\amp}{\langle i\Gamma_{em}\rangle}

\def\gappeq{\mathrel{\rlap {\raise.5ex\hbox{$>$}}
{\lower.5ex\hbox{$\sim$}}}}

\def\lappeq{\mathrel{\rlap{\raise.5ex\hbox{$<$}}
{\lower.5ex\hbox{$\sim$}}}}

\def\Toprel#1\over#2{\mathrel{\mathop{#2}\limits^{#1}}}

\begin{document}

\title{Coulomb corrections to inclusive cross sections at the future Electron - Ion Collider}
\author{V.P. Gon\c{c}alves$^1$, M.V.T Machado$^2$, F.S. Navarra$^3$ and  D. Spiering$^3$}
\affiliation{$^1$Instituto de F\'{\i}sica e Matem\'atica,  Universidade
Federal de Pelotas,
C.P. 354, CEP 96010-900, Pelotas, RS, Brazil\\
$^2$High Energy Physics Phenomenology Group, GFPAE  IF-UFRGS \\
Caixa Postal 15051, CEP 91501-970, Porto Alegre, RS, Brazil\\
$^3$Instituto de F\'{\i}sica, Universidade de S\~{a}o Paulo, 
C.P. 66318,  05315-970, S\~{a}o Paulo, SP, Brazil\\
}
\begin{abstract}
The experimental results of the future electron -- ion ($e A$) collider are expected to constrain the dynamics of the strong interactions 
at small values of the Bjorken -- $x$ variable and large nuclei. Recently it has been suggested that  Coulomb corrections can be 
important in inclusive and diffractive $eA$ interactions. In this paper we present a detailed investigation of the impact of the Coulomb 
corrections to some of the observables  that will be measured in the future $eA$ collider. In particular, we estimate the magnitude of these 
corrections for the charm and longitudinal cross sections in inclusive and diffractive interactions. Our results demonstrate that the Coulomb 
corrections for these observables are negligible, which implies that they can be used to probe the QCD dynamics. 
\end{abstract}
\maketitle

\section{Introduction}

The future Electron - Ion Collider (EIC) \cite{Raju,Boer,Accardi,raju2} will allow us to study the hadronic structure in the regime of large 
partonic densities and high strong field  strengths, which are expected to modify the linear evolution equations \cite{dglap,bfkl} normally used in  
Quantum Chromodynamics (QCD) at high energies. We expect that the EIC experimental data will be able to 
establish the presence of gluon saturation effects, the magnitude of the associated non-linear corrections and determine what is the correct theoretical 
framework for their description  (For reviews see, e.g., Ref. \cite{hdqcd}). In particular, the enhancement of the non-linear effects with the 
nuclear mass number  has motivated the development of several studies on the implications of these effects in inclusive and  exclusive 
observables which could be measured in  Electron - Ion Colliders \cite{victor,simone1,simone2,Nik_schafer,Kowalski_prl,Kowalski_prc,erike_ea1,erike_ea2,vmprc,simone_hq,
Caldwell,vic_erike,Lappi_inc,Toll,joao,Lappi_bal,diego,diegojpg}. These studies have demonstrated that the $eA$ collider is the ideal environment 
to improve our undestanding of the strong interactions at high energies. 

Deep inelastic $eA$ scattering is characterized by a large electron energy
loss $\nu$ (in the target rest frame) and an invariant momentum
transfer $q^2 \equiv - Q^2$ between the incoming and outgoing
electron such that $x = Q^2/2m_N \nu$ is fixed ($m_N$ is the target mass). 
In terms of Fock states we then view  $eA$ scattering as follows: the
electron emits a photon ($|e\!> \rightarrow |e\gamma\!>$) with
$E_{\gamma} = \nu$ and $p_{t \, \gamma}^2 \approx Q^2$. Afterwards  the
photon splits into a $q \overline{q}$ pair ($|e\gamma\!> \rightarrow |e
q\overline{q}\!>$), which has electric and color charges, and typically travels a distance $l_c \approx
1/m_N x$, referred to as the coherence length, before interacting in
the target. For small $x$, the photon is converted into  a $q \bar{q}$
pair at a large distance before the scattering.
Consequently, the space-time picture of the DIS in the target rest
frame can be viewed as the decay of the virtual photon at high
energy  into a quark-antiquark pair, which subsequently interacts with the 
target. If the target is a large -- $A$  nucleus, it is characterized by a strong color field, generated by the fusion of the color 
fields of its nucleons. Consequently, in order to describe the  dipole -- nucleus scattering one must take into account the multiple 
interactions of the dipole  with the nuclear color field \cite{hdqcd}. This aspect has been considered in the phenomenological works that have 
studied the impact of the non - linear corrections to the QCD dynamics on the observables that will be measured in the future $eA$ collider 
\cite{victor,simone1,simone2,Nik_schafer,Kowalski_prl,Kowalski_prc,erike_ea1,erike_ea2,vmprc,simone_hq,
Caldwell,vic_erike,Lappi_inc,Toll,joao,Lappi_bal,diego,diegojpg}. However, these studies have not taken into account the contribution of the 
Coulomb corrections to the dipole -- nucleus interaction. As the $q \bar{q}$ dipole has electric charge and a large nucleus generates a strong 
electromagnetic  field, we might  expect that the nuclear DIS  receives a substantial contribution from electromagnetic interactions of the 
dipole with the nucleus, which is known as Coulomb correction. A first analysis of the impact of these corrections to the DIS off a heavy nucleus 
has been performed in Ref. \cite{TuchinPRL} (See also Ref. \cite{boris}) and has indicated that  the Coulomb corrections reach to 20 \% of the 
inclusive cross section and about 40\% of the diffractive cross section in the kinematical range of small values of the Bjorken-$x$ variable and 
low photon virtualities $Q^2$. Considering the large magnitude predicted for these corrections, it is important to investigate in more detail how 
the behavior of different observables will be modified by the Coulomb corrections in the kinematical range that will be probed in the future EIC.  
Apart from the total inclusive and diffractive cross sections discussed in Ref. \cite{TuchinPRL}, the EIC is expected to measure the heavy quark 
and longitudinal contributions to these cross sections. In this paper we extend the study perfomed in Ref. \cite{TuchinPRL} and analyze the impact 
of the Coulomb corrections on these quantities. Our goal is to verify if it is possible to disentangle the Coulomb corrections from the QCD interactions 
in these different observables and, consequently,  to probe the QCD dynamics at high energies.

This paper is organized as follows. In the  next Section we briefly review the description of the photon-nucleus interaction in the dipole approach 
and establish the relation between the measured inclusive and diffractive cross sections and the color/electric dipole target elastic scattering amplitude. 
Moreover, the treatment of the strong and electromagnetic interactions in the nuclear field is presented.  In Section \ref{results} 
we present a detailed study of the Coulomb corrections to the  charm  and longitudinal contributions to the inclusive and diffractive cross sections. 
Our focus is on  predictions for DIS off a heavy nucleus in the kinematical region of small values of $x$ and  $Q^2$. Finally, in Section \ref{conc} we 
summarize our main  results and conclusions.

\section{Theoretical framework}

The basic idea of the dipole approach is that the  electron-nucleus interaction at high energies can be factorized in terms of the fluctuation of 
the virtual photon into a $q \bar{q}$ dipole and the dipole-nucleus scattering. As a consequence, the longitudinal and transverse $\gamma^*A$ cross 
sections can be expressed as follows \cite{dipole},
\begin{eqnarray}
\sigma_{L,T}^{inc}(x,Q^2)& = & \sum_f \int dz \,d^2\rr
|\Psi_{L,T}^{(f)}(z,\rr,Q^2)|^2
\,2 \int d^2 \rb \, {\cal{N}}(x,\rr,\rb)\,\,,
\label{sigma_inc}
\end{eqnarray}
where the  variable $\rr$ defines the relative transverse
separation of the pair (dipole) and $z$ $(1-z)$ is the
longitudinal momentum fraction of the quark (antiquark) of flavor $f$. In the case of diffractive $eA$ collisions, the dipole approach predicts that 
the cross sections can be expressed as follows
\begin{eqnarray}
\sigma_{L,T}^{\mathrm{diff}}(x,Q^2)& = & \sum_f \int dz d^2\rr
|\Psi_{L,T}^{(f)}(z,\rr,Q^2)|^2
 \,\int d^2 \rb \,\left|{\cal{N}}(x,\rr,\rb)\right|^2\,\,.
\label{sigma_diffrac}
\end{eqnarray}
  The
photon wave functions $\Psi_{L,T}$ are determined from light cone
perturbation theory and are given by  
\begin{eqnarray}
|\Psi_{T} (z,\rr,\,Q^2)|^2  =  \frac{6\alpha_{\mathrm{em}}}{4\,\pi^2}  \, 
\sum_f e_f^2
 \, {[z^2 + (1-z)^2]\, \varepsilon^2 \, K_1^2(\varepsilon \,\rrn)
 + m_f^2 \, \, K_0^2(\varepsilon\, \rrn)} & &
\label{wtrans}
\end{eqnarray}
and
\begin{eqnarray}
|\Psi_{L} (z,\rr,\,Q^2)|^2  =  \frac{6\alpha_{\mathrm{em}}}{\pi^2} \,
\sum_f e_f^2 
\, \left\{Q^2 \,z^2 (1-z)^2 \, K_0^2(\varepsilon\, \rrn) \right\} .  & &
\label{wlongs}
\end{eqnarray}
The
auxiliary variable $\varepsilon^2=z(1-z)\,Q^2 + m^2_f$ depends on
the quark mass, $m_f$. The $K_{0,1}$ are the Mcdonald functions
and the summation is performed over the quark flavors. Here, we consider the light quark masses to be $m_f = 0.14$ GeV (for $f=u,d,s$) and for the 
heavy ones we consider $m_c = 1.4$ GeV and $m_b=4.5$ GeV.

The function ${\cal{N}}(x,\rr,\rb)$ is the  quark dipole-nucleus forward scattering amplitude for a given impact parameter $\rb$  which encodes all 
the information about the scattering due to strong and electromagnetic interactions. In order to estimate this quantity we will consider the  
Glauber multiple scattering approach \cite{glauber,glaubermat}, which is justified at high energies when the incident particle passes through the 
nucleus in a very short time, so that the change of the positions of the nucleons can be neglected. In this model the scattering amplitude obtained by 
calculating the scattering from instantaneously fixed nucleons is  then averaged over the initial and final nucleus wave functions. The main assumption 
is that the total eikonal phase acquired by the particle when passing through the nucleus at a fixed impact parameter is equal to the sum of the phases 
from the individual nucleons. During the last decades, the Glauber model has been sucessfully applied to describe proton - nucleus collisions 
(For a review see e.g. Ref. \cite{alkhazov}). In particular, the impact of the Coulomb corrections has been studied by several authors (See e.g. 
Refs. \cite{glaubermat,czyz,lesniak,bleszynski}), who have demonstrated that it plays an important role in heavy nuclei, influencing the differential 
cross sections around the minima of the squared momentum transfer ($t$)  distributions. The basic assumption in these studies is that the proton - 
nucleon phase shift can be described by the sum of purely strong and Coulomb phase shifts \cite{glaubermat,czyz,lesniak}. In 
Refs. \cite{TuchinPRD,TuchinPRL,TuchinPRC},  the Glauber model has been applied to describe the dipole - nucleus interaction and the additivity of 
the strong and Coulomb phase shifts to estimate the dipole - nucleon was assumed. As a consequence, the dipole - nucleus  
forward scattering amplitude can be expressed as follows \cite{TuchinPRL,TuchinPRC}
\begin{eqnarray}
  \N (x,\rr,\rb) =  1-\cos\left[Z\langle\real i\Gamma_{em}\rangle\right]\exp\left[-A\langle\imaginary i\Gamma_s\rangle\right]\,,
  \label{sec_dip}
\end{eqnarray}
where $i\Gamma_{em}$ and $i\Gamma_s$ are the electromagnetic and strong contribution to the dipole - nucleon elastic scattering amplitude, respectively. 
At large impact parameters ($b > R_A$) the strong interaction contribution is expected to be small, i.e. $A[i\Gamma_s]\rightarrow0$, and hence
\begin{eqnarray}
  \N (x,\rr,\rb) \approx \N_{em}(x,\rr,\rb) = 1 - \cos\left[Z\langle i\Gamma_{em}\rangle\right]\,.
  \label{N_em}
\end{eqnarray}
On the other hand, at small impact parameters ($b < R_A$), the strong interaction becomes dominant and the dipole - nucleus scattering amplitude can be 
approximated by 
\begin{eqnarray}
  \N (x,\rr,\rb) \approx 
  \N_s(x,\rr,\rb) = 1 - \exp\left[-A\langle i\Gamma_s\rangle\right]\,.
\end{eqnarray}
In what follows we will assume that these asymptotic solutions can be used to estimate the  dipole -- nucleus scattering amplitude for $b \sim R_A$, which is the region of interest for the deep inelastic scattering.  In order to calculate the inclusive and diffractive cross sections we must  specify the models used for the electromagnetic and strong dipole -- nucleus 
scattering amplitude. Following Ref. \cite{TuchinPRL} we will assume that 
\begin{eqnarray}
 {\cal{N}}_{s}(x,\rr,\rb)  =  
\left[\, 1- \exp \left(-\frac{ Q_{\mathrm{s},A}^2(x)\,\rr^2}{4} \right) \, \right] \Theta \left(R_A-b \right),
  \label{N_GBW}
\end{eqnarray}
where $ Q_{\mathrm{s},A}$  is the  nuclear saturation scale, which we will assume to be given by $Q_{\mathrm{s},A}^2(x) = A^{1/3}Q_{\mathrm{s}}^2 $, 
where 
$Q_{\mathrm{s}}^2 (x) = Q_0^2\,e^{\lambda\ln(x_0/x)}$ is the saturation scale for a proton. This model is a naive generalization for the nucleus of the 
saturation model 
proposed in Refs. \cite{GBW}, which captures the main properties of the high energy evolution equations and it is suitable for describing the non -- linear 
physics in the small-$x$ region. In particular, this model implies that ${\cal{N}}_{s} \propto r^2$ (color transparency) at small pair separations and 
that the multiple scatterings are ressumed in a Glauber -- inspired way.
  Such model was used in Refs. \cite{simone1,simone2} to estimate the impact of the saturation physics in the observables that will be measured in the 
future $eA$ collider. In our calculations we will consider the parameters obtained from a fit to the HERA data in Ref. \cite{GBW}:  $Q_0^2 = 1$ GeV$^2$,
 $\lambda= 0.277$ and $x_0=3.41 \cdot 10^{-4}$. 
Using Eq. (\ref{N_GBW}) we obtain that 
\begin{eqnarray}
 2 \int_{b<R_A} d^2\rb \, \N(x,\rr,\rb) \approx  2 \int_{b<R_A} d^2\rb \, \N_{s}(x,\rr,\rb) =  2\pi R_A^2 \left(1 - e^{-Q_{\mathrm{s},A}^2r^2/4}\right)\,.
\label{Nsinc_inc}
\end{eqnarray}
and 
\begin{eqnarray}
 \int_{b<R_A} d^2\rb \left|\N(x,\rr,\rb)\right|^2  & \approx & \int_{b <R_A} d^2\rb \left|\N_s(x,\rr,\rb)\right|^2 
=  \pi R_A^2 \left(1 - e^{-Q_s^2r^2/4}\right)^2\,.
\label{Nsinc_dif}
\end{eqnarray}
For the electromagnetic case we will follow the approach proposed in Refs. \cite{TuchinPRL,TuchinPRC}. 
In particular, we will assume that the leading term in $\alpha$ for the dipole - nucleon scattering arises from a single photon exchange, with the elastic dipole - nucleon amplitude being approximately real.  As demonstrated in Ref. \cite{TuchinPRD}, at the Born approximation, the contribution of the imaginary part is proportional to $\alpha^2 Z \sim \alpha$ in the $Z\alpha \sim 1$, while the contribution of the real part is of the order of $\alpha^2 Z^2 \sim 1$. Considering the Weizscker - Williams approximation, where the $t$ - channels photons at high energies are assumed to be almost real,  the electromagnetic dipole - nucleon elastic scattering amplitude  was derived in Ref.\cite{TuchinPRD}, being given by
\begin{eqnarray}
  \amp \approx 2\alpha\ln\left(\frac{\left| \rb-\rr/2 \right|}{\left| \rb + \rr/2\right|}\right)
  = \alpha \ln\left(\frac{b^2+r^2-br\cos\phi}{b^2+r^2+br\cos\phi}\right)\,.
  \label{ampelet}
\end{eqnarray}
Moreover, we will take into account that for the deep inelastic scattering the main contributions for the DIS cross sections comes from values of pair separations of the order  $r \sim 1/m_f$ \cite{simone1,simone2}. Consequently, we will have that in general $r \ll b$, which allows to simplify the expression for the dipole - nucleon scattering amplitude. Such inequality becomes stronger in the case of the heavy quarks. In particular, for  $b\gg r$ the Eq. (\ref{ampelet}) can be simplified and becomes
\begin{eqnarray}
  \amp = 2\alpha\frac{\rb\cdot\rr}{b^2} + \mathcal{O}(r^3/b^3)
   \approx \frac{2\alpha r\cos\phi}{b}\,,
   \label{b>r}
\end{eqnarray}
which implies that
\begin{eqnarray}
  \cos\left[Z\amp\right] = \cos\left( \frac{2\alpha Zr\cos\phi}{b} \right)
  = 1 - \frac{1}{2}\left(\frac{2\alpha Zr\cos\phi}{b}\right)^2 + \mathcal{O}(r^4/b^4)
  \label{cos_exp}\approx 1 -\frac{2\alpha^2Z^2 r^2\cos^2\phi}{b^2}\,,
\label{cos_approx}
\end{eqnarray} 
and 
\begin{eqnarray}
  \N_{em}(x,\rr,\rb) \approx \frac{2\alpha^2Z^2r^2\cos^2\phi}{b^2}\,.
  \label{N_s_final}
\end{eqnarray}
Consequently, we have
\begin{eqnarray}
 2 \int_{b>R_A} d^2\rb \, \N(x,\rr,\rb) \approx  2 \int_{b>R_A} d^2\rb \, \N_{em}(x,\rr,\rb)
    & = & 4\alpha^2Z^2r^2\int_{R_A}^{b_{max}}\frac{db}{b}\int_0^{2\pi}d\phi \cos^2\phi\\
   &=& 4\pi\alpha^2Z^2r^2\ln\left(\frac{b_{max}}{R_A}\right)
   = 4\pi\alpha^2Z^2r^2\ln\left(\frac{W^2}{4m_f^2m_NR_A}\right)\,,
\label{Ninc_em_int}
\end{eqnarray}
where $m_N$ is the nucleon mass and $W$ is the photon - nucleus center of mass energy.
Moreover, $b_{max} = s/(4m_N m_f^2)$ is a long distance cutoff of the $b$ integral, which is directly related to the minimum transverse momentum at which the Weizsacker - Williams approximation still is valid  \cite{TuchinPRD}. As in demonstrated in Ref. 
\cite{TuchinPRD}, one have that $b_{max} \gg R_A$, which justifies the approximations used in the derivation of the Eqs. (\ref{ampelet}) -- (\ref{Ninc_em_int}).
Considering the same approximations,  for $b > R_A$ we find  
\begin{eqnarray}
  \left|\N(x,\rr,\rb)\right|^2 \approx \left|\N_{em}(x,\rr,\rb)\right|^2 = \left| 1-\cos\left[Z\langle\real i\Gamma_{em}\rangle\right]\right|^2 
= 1 - 2\cos\left[Z\langle\real i\Gamma_{em}\rangle\right] + \cos^2\left[Z\langle\real i\Gamma_{em}\rangle\right]\,.
  \label{N2cos}
\end{eqnarray}
Using that $\cos^2\left[Z\amp\right] = 1 + \mathcal{O}(\alpha^4 r^4/b^4)$, yields
\begin{eqnarray}
  \left|\N_{em}(x,\rr,\rb)\right|^2  \approx \frac{4\alpha^2Z^2r^2\cos^2\phi}{b^2}\,,
  \label{N2_diff}
\end{eqnarray}
which implies that 
\begin{eqnarray}
 \int_{b>R_A} d^2\rb \left|\N(x,\rr,\rb)\right|^2  & \approx & \int_{b>R_A} d^2\rb \left|\N_{em}(x,\rr,\rb)\right|^2  
= 4\pi\alpha^2Z^2r^2\ln\left(\frac{W^2}{4m_f^2m_NR_A}\right)\,.
 \label{N2_em_int}
\end{eqnarray}
As obtained in Ref. \cite{TuchinPRL},  the electromagnetic contribution for the inclusive and diffractive cross sections are equal, differently 
from the strong one. As a consequence we expect that its impact on inclusive and diffractive observables will be different. In the next Section 
we will estimate how the electromagnetic contribution modifies these observables in the kinematical range that will be probed by the future 
electron - ion collider.

\begin{figure*}[t]
\includegraphics[scale=0.35]{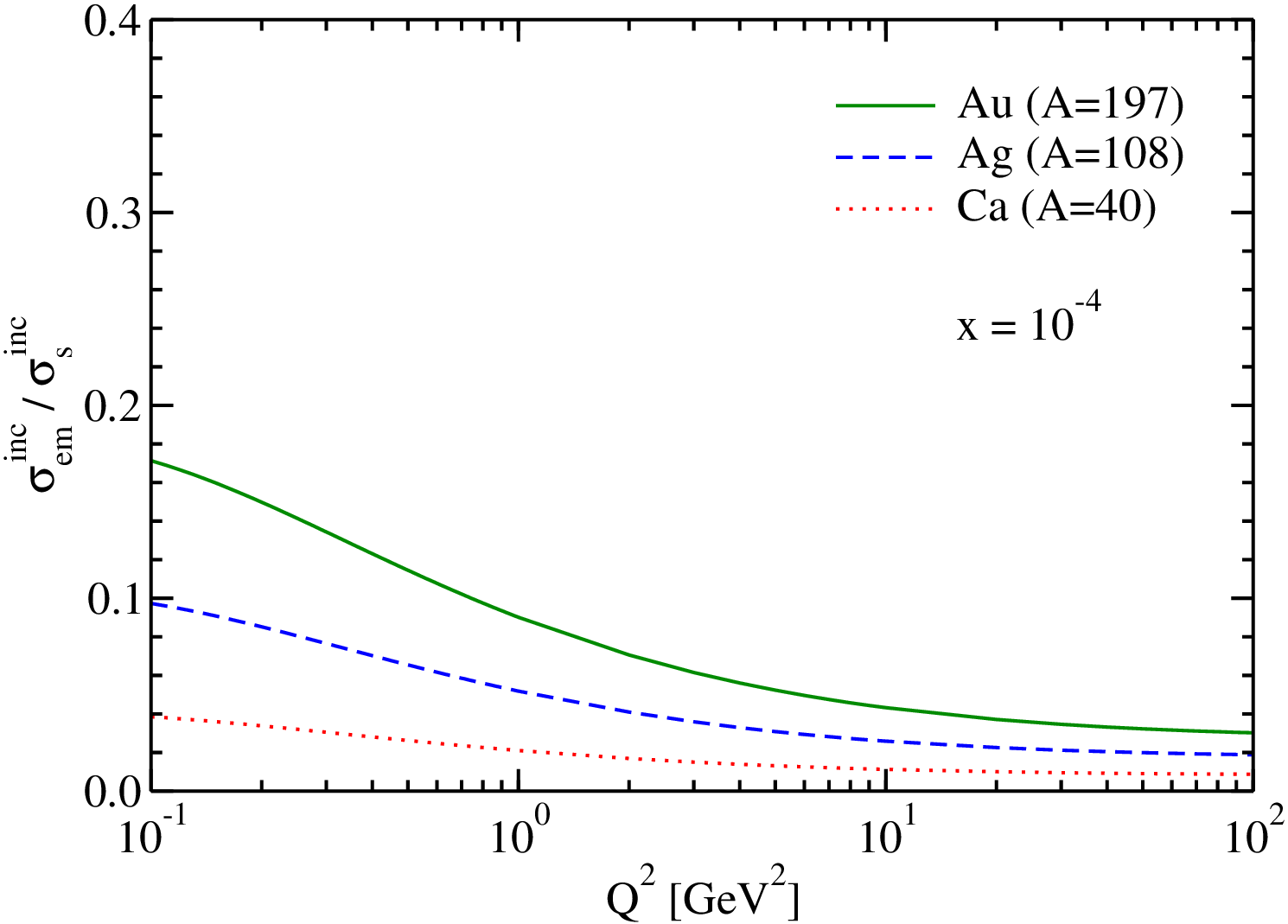} 
\includegraphics[scale=0.35]{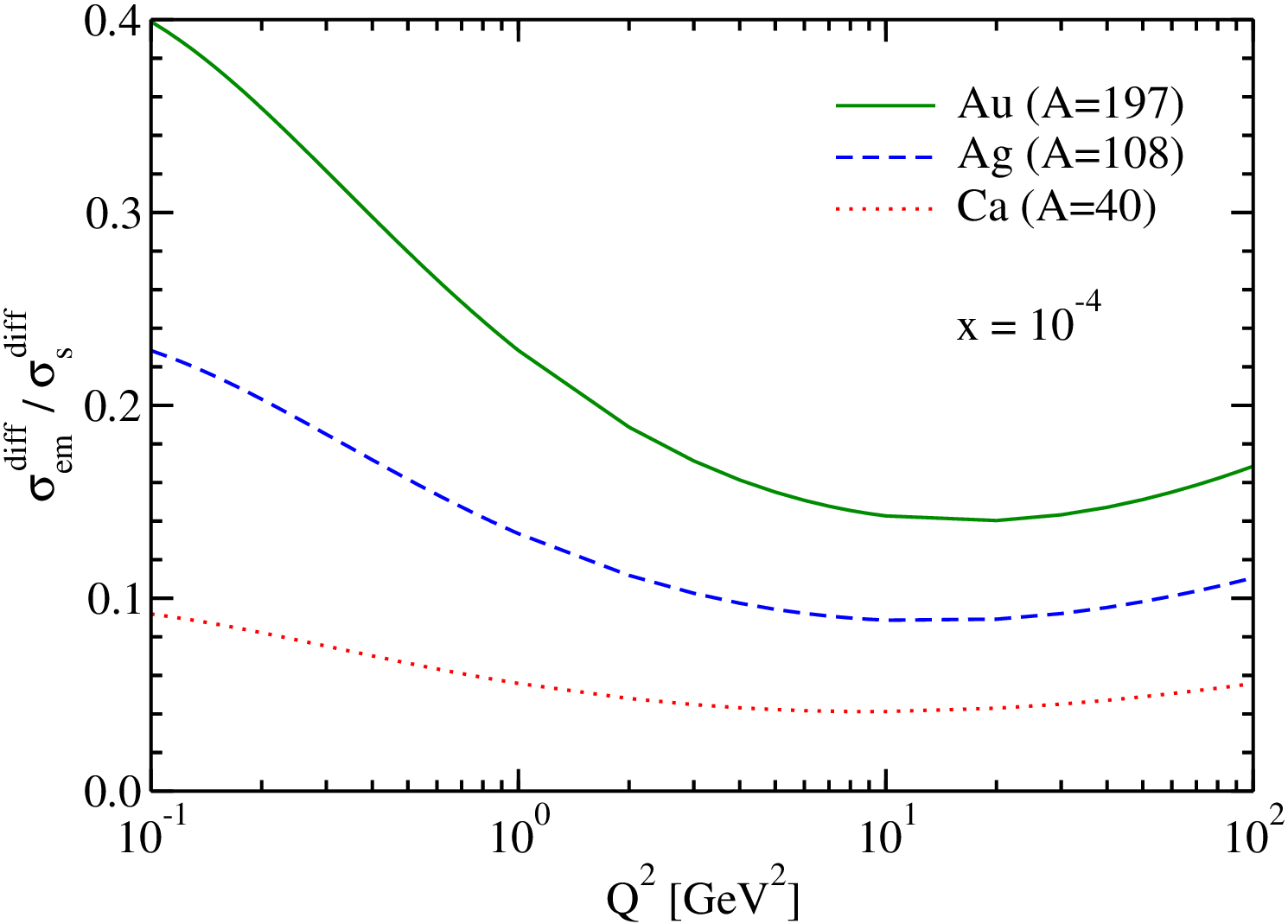}
 \caption{Dependence on the photon virtuality $Q^2$ for different atomic numbers and $x= 10^{-4}$ of the ratio between the electromagnetic and 
strong contributions for the inclusive (left panel) and diffractive (right panel) cross sections.}
\label{fig1}
\end{figure*}

\begin{figure*}[t]
\includegraphics[scale=0.35]{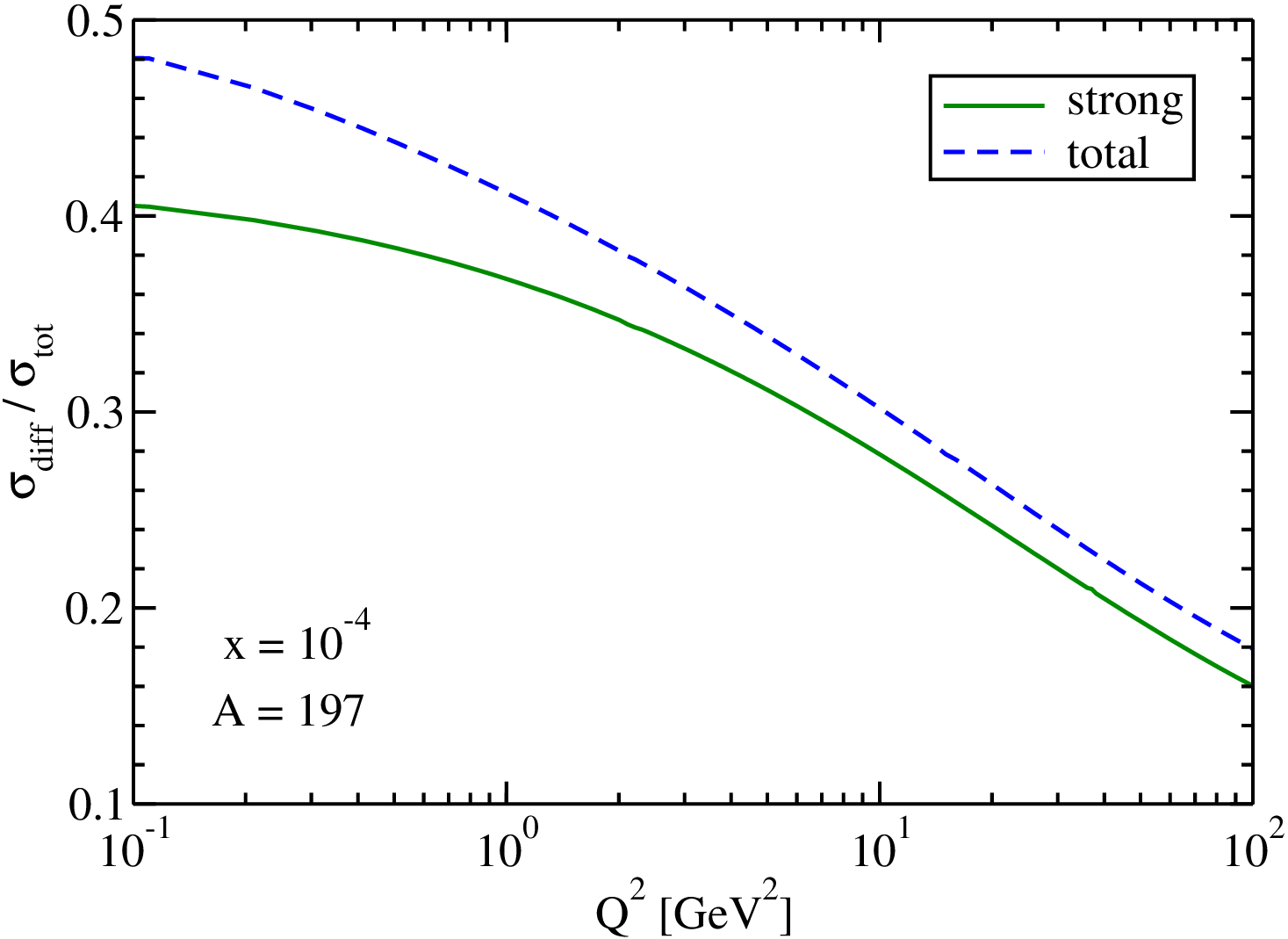} 
\includegraphics[scale=0.35]{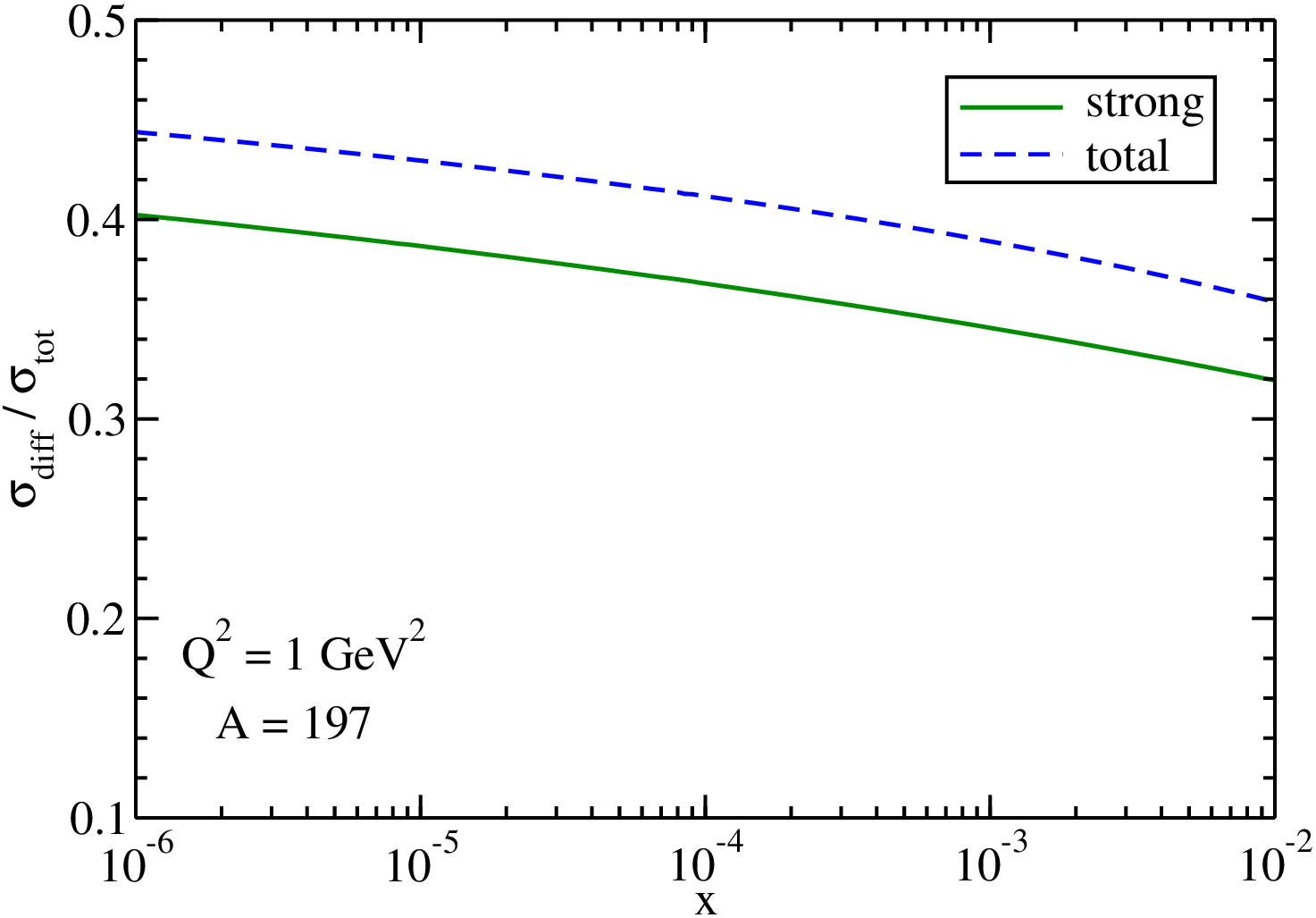}
 \caption{Ratio $\sigma_{diff}/\sigma_{tot}$ as a function of $Q^2$ for fixed $x$ (left panel) and as a function of  $x$ for fixed $Q^2$ (right panel).}
\label{fig1b}
\end{figure*}

\begin{figure*}[t]
\includegraphics[scale=0.35]{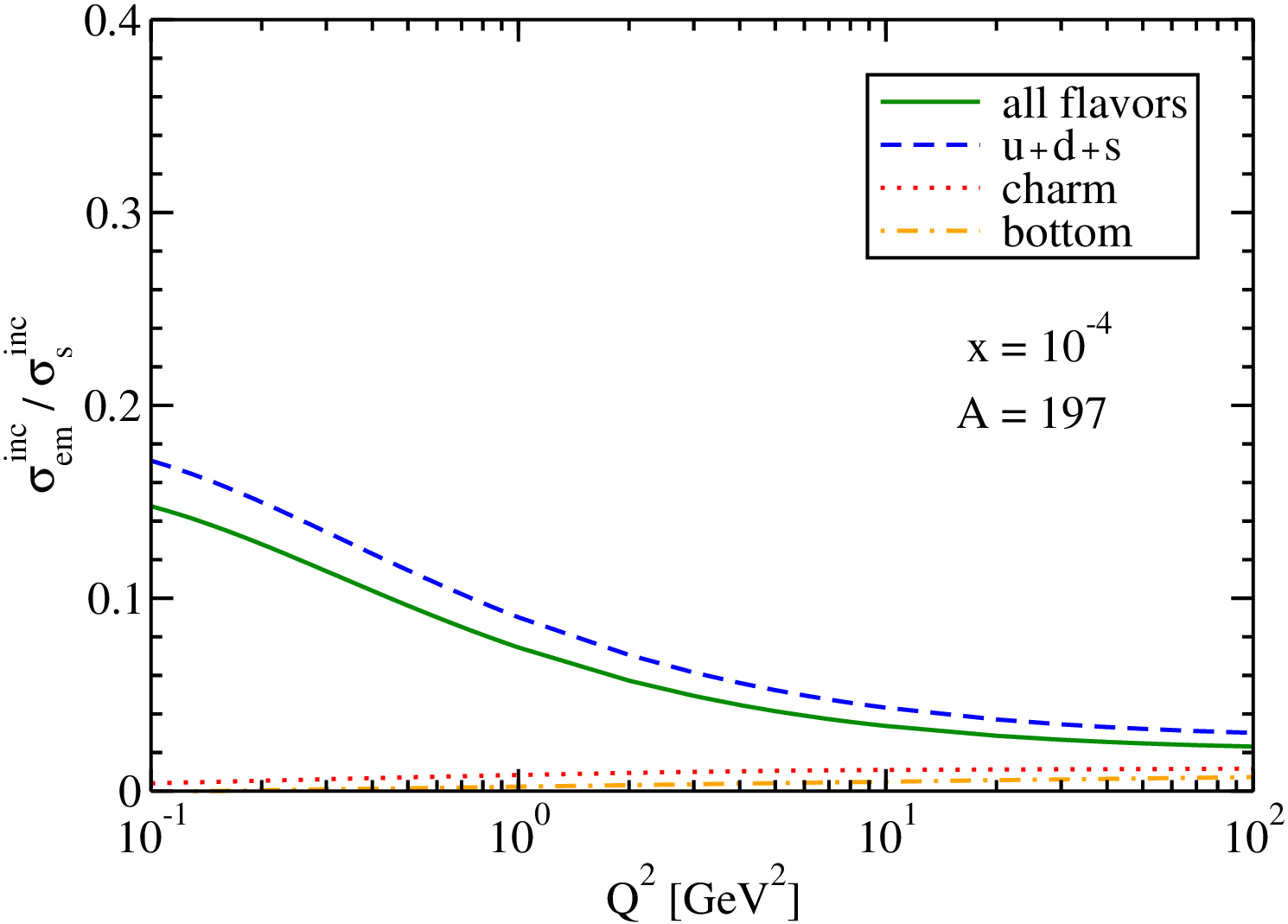} 
\includegraphics[scale=0.35]{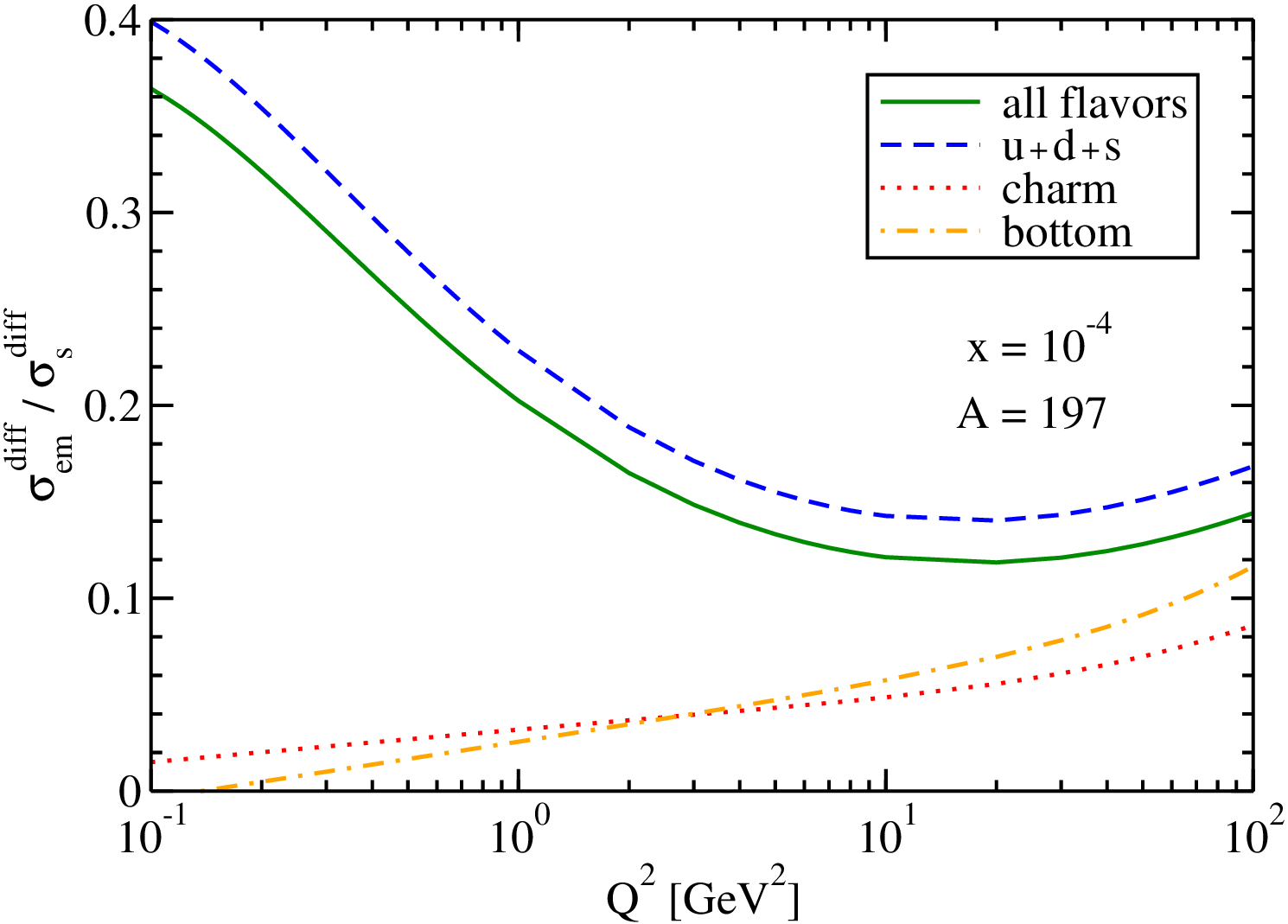}
 \caption{Flavor decomposition for the ratio between the electromagnetic and strong contributions for the inclusive (left panel) and diffractive 
(right panel) processes  as a function of photon virtuality $Q^2$ and fixed $x=10^{-4}$ for gold nucleus.}
\label{fig2}
\end{figure*}

\begin{figure}[t]
  \centering
  \includegraphics[scale=0.35]{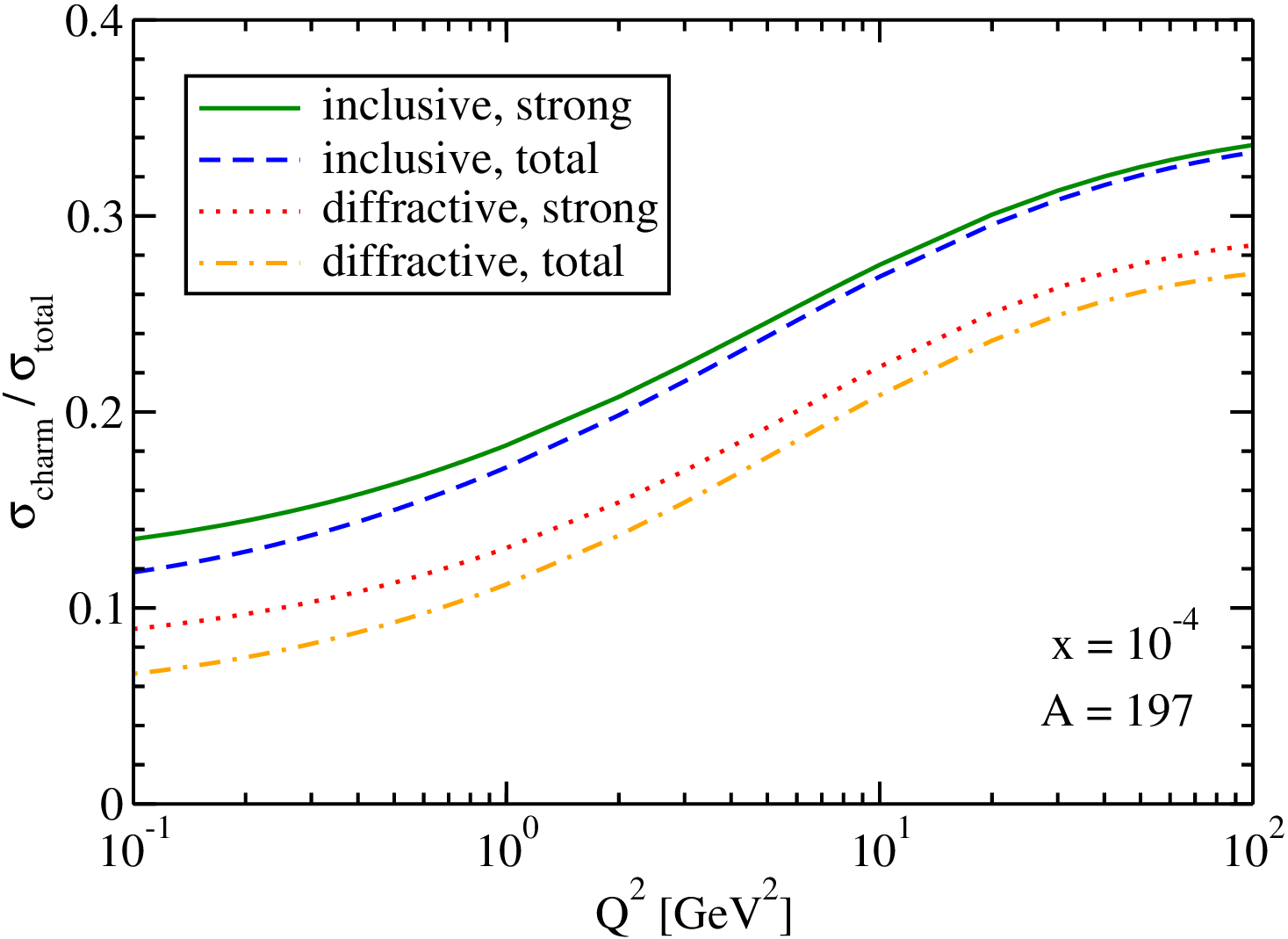}
  \includegraphics[scale=0.35]{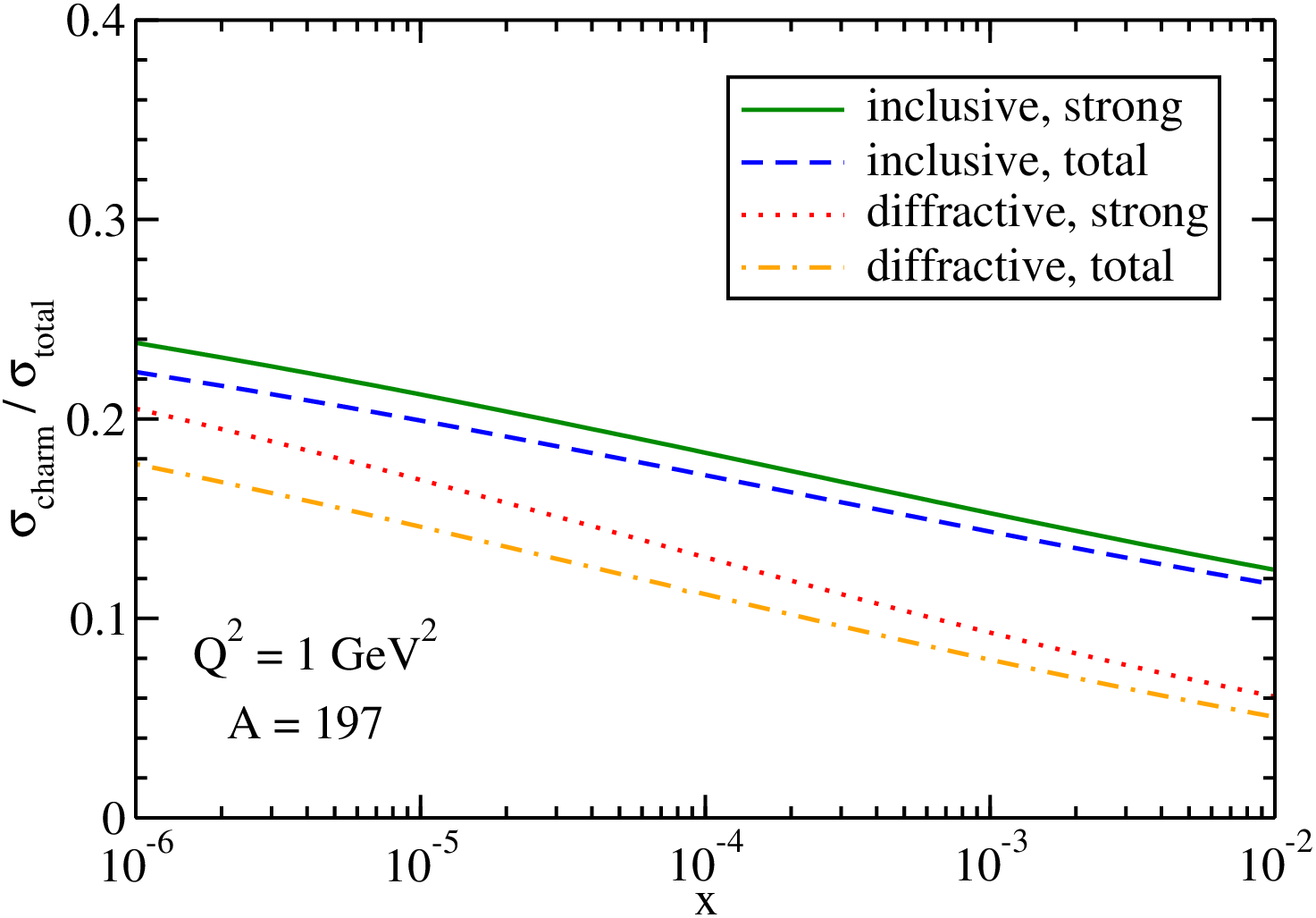}  
 \caption{Dependence on $Q^2$ (left panel) and $x$ (right panel) of the ratios between the charm and total cross sections for inclusive and 
diffractive interactions. }
\label{fig3}
\end{figure}

\section{Results and discussion}
\label{results}

In this Section we present a detailed study of the flavor and polarization decomposition of inclusive and diffractive cross sections including the Coulomb 
correction. Improving the work done in  Ref. \cite{TuchinPRL}, we will include in our analysis the contribution of the heavy quarks (charm and bottom). 
As discussed e.g. in Ref. \cite{simone1,erike_ea2}, and shown in what follows, the charm contribution is expected to be $\approx 20 \%$ of the total cross 
section at small values of $x$ and low virtualities. Moreover, it is considered an important probe of the nuclear gluon distribution. Similar expectation also 
motivates the analysis of the longitudinal cross section (See e.g. \cite{erike_ea1,zurita}).  Our goal is to verify the impact of the Coulomb corrections 
in these observables in inclusive and diffractive interactions.  

We start our analysis presenting in Fig \ref{fig1} the ratio between the  Coulomb  and strong (QCD) contributions to the inclusive (left panel) and 
diffractive (right panel) cross sections.  In Fig. \ref{fig1} the ratios are shown as a function of photon virtuality $Q^2$ for a fixed value of Bjorken 
variable, $x=10^{-4}$, and three distinct nuclei: Gold (solid line), Silver (dashed line) and Calcium (dotted line). 
The results are consistent with those obtained in Ref. \cite{TuchinPRL}, even including the heavy quark contribution, showing  that the Coulomb correction  
is quite important in the kinematical region of low-$Q^2$ and small-$x$ and it is enhanced for the heavier nuclei. In the perturbative range  that will be
probed in the future EIC ($Q^2 \gtrsim 1$ GeV$^2$) and $A = 197$,  the Coulomb contribution is $\approx 10 \, (21) \%$ for the inclusive 
(diffractive) cross section. The different impact of the  Coulomb interactions on the inclusive and diffractive cross sections can be understood as follows. 
In contrast to the inclusive cross section, the main contribution to the diffractive cross section comes from larger dipoles (See e.g. Ref. \cite{simone2}) and  
the electromagnetic terms are proportional to $r^2$ [See Eqs. (\ref{Ninc_em_int}) and (\ref{N2_em_int})].  
As a consequence, the presence of  Coulomb corrections  modifies  the ratio between the diffractive and the total cross section, which is 
expected to be measured in the future electron - ion collider. Our predictions for the $Q^2$ and $x$ dependence of this ratio are presented in 
Fig. \ref{fig1b}. We compare the predictions obtained with the sum of the electromagnetic and strong contributions, denoted total in the figure, with 
those derived disregarding the Coulomb corrections. We observe that the ratio is enhanced by the electromagnetic contribution, in particular at 
low $Q^2$ and small -- $x$. At $Q^2 = 1$ GeV$^2$, this enhancement is $\approx 10 \%$.

The $r^2$ dependence of the electromagnetic contribution has a direct impact on the Coulomb corrections to the heavy quark  cross sections. 
As these cross sections  are dominated by small dipoles, we expect that the effect of the Coulomb corrections will be smaller here than  in the    
case of  light quark production. This expectation is confirmed by the results  shown in Fig. \ref{fig2}, where we present the flavor decomposition 
of the ratio between the electromagnetic and strong contributions.  We observe that for $Q^2 \approx 1$ GeV$^2$ the Coulomb corrections can be 
disregarded in the charm and bottom production in the inclusive case and are of order of 2 \% in diffractive interactions. At larger values of $Q^2$, 
the heavy quark contribution to the total cross sections increases as well as the Coulomb corrections. We have verified that the impact of the Coulomb 
corrections to the charm and bottom production is almost $x$ independent in the range 
$1 \le Q^2 \le 10$ GeV$^2$. Another interesting aspect is that the inclusion of the heavy quarks decreases the magnitude of the Coulomb corrections 
for the total cross sections in comparison to those obtained considering only light quarks, denoted by $u + d + s$ in the figure.

\begin{figure}[t]
  \centering
  \includegraphics[scale=0.33]{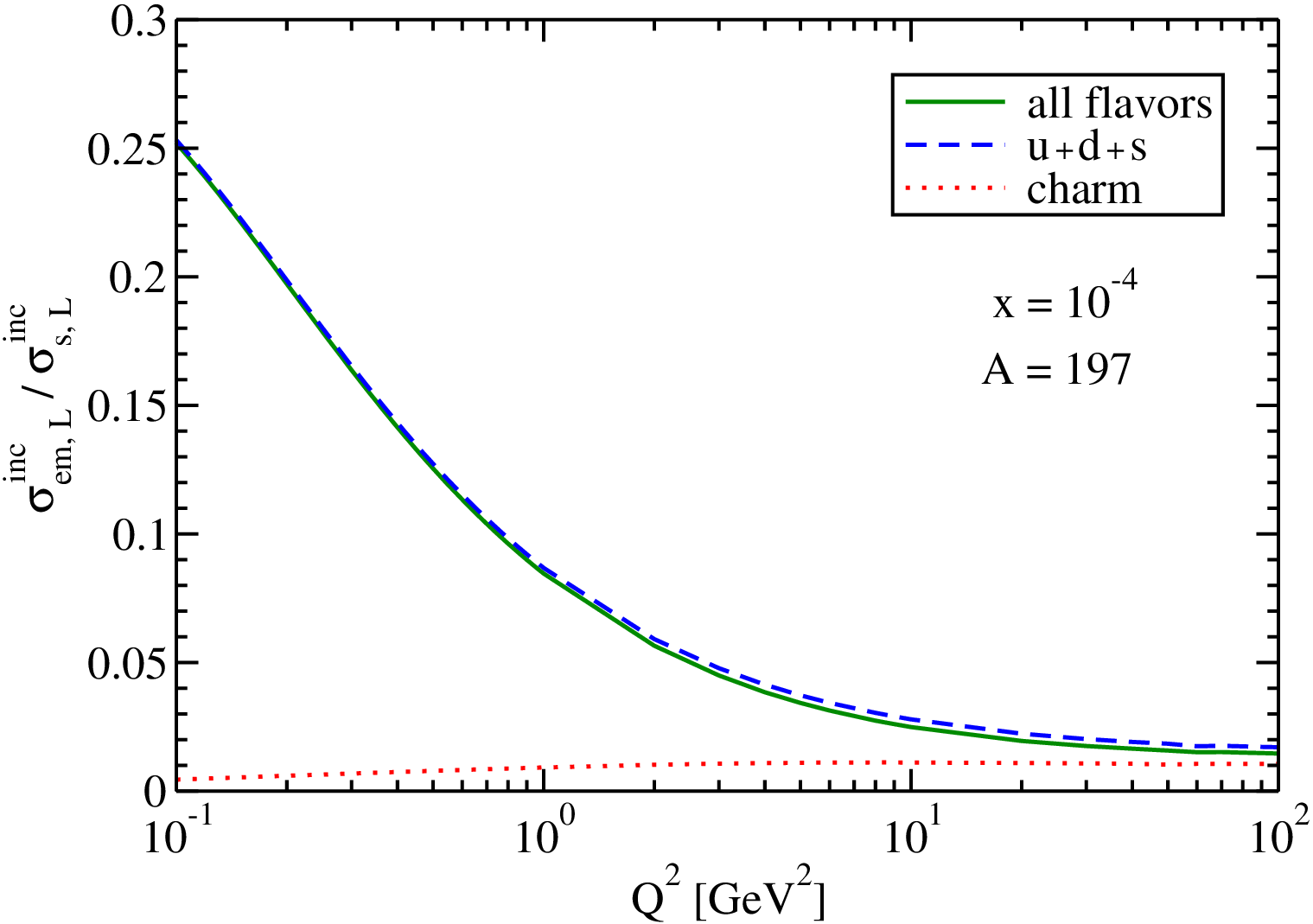}
  \includegraphics[scale=0.33]{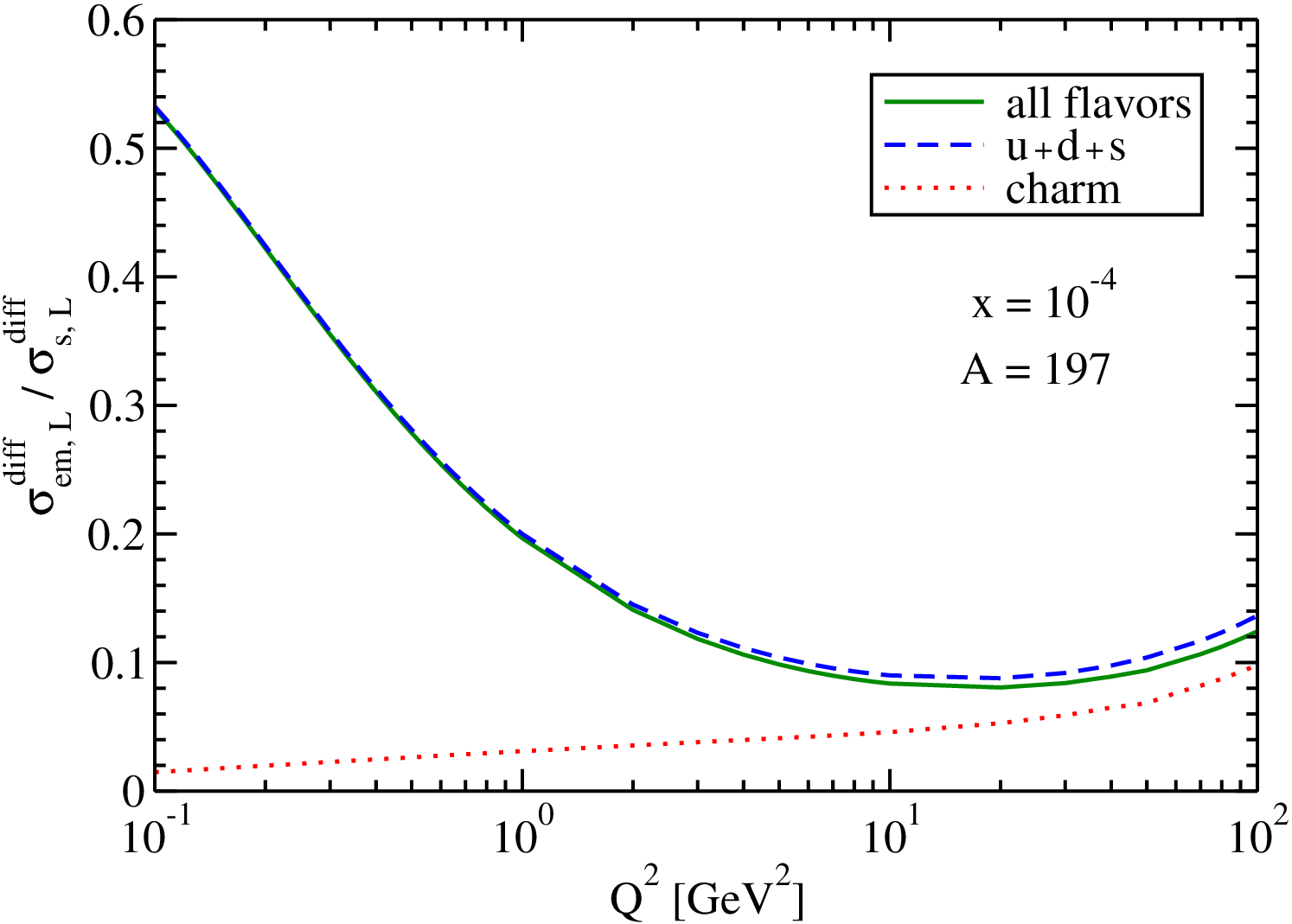}
  \includegraphics[scale=0.33]{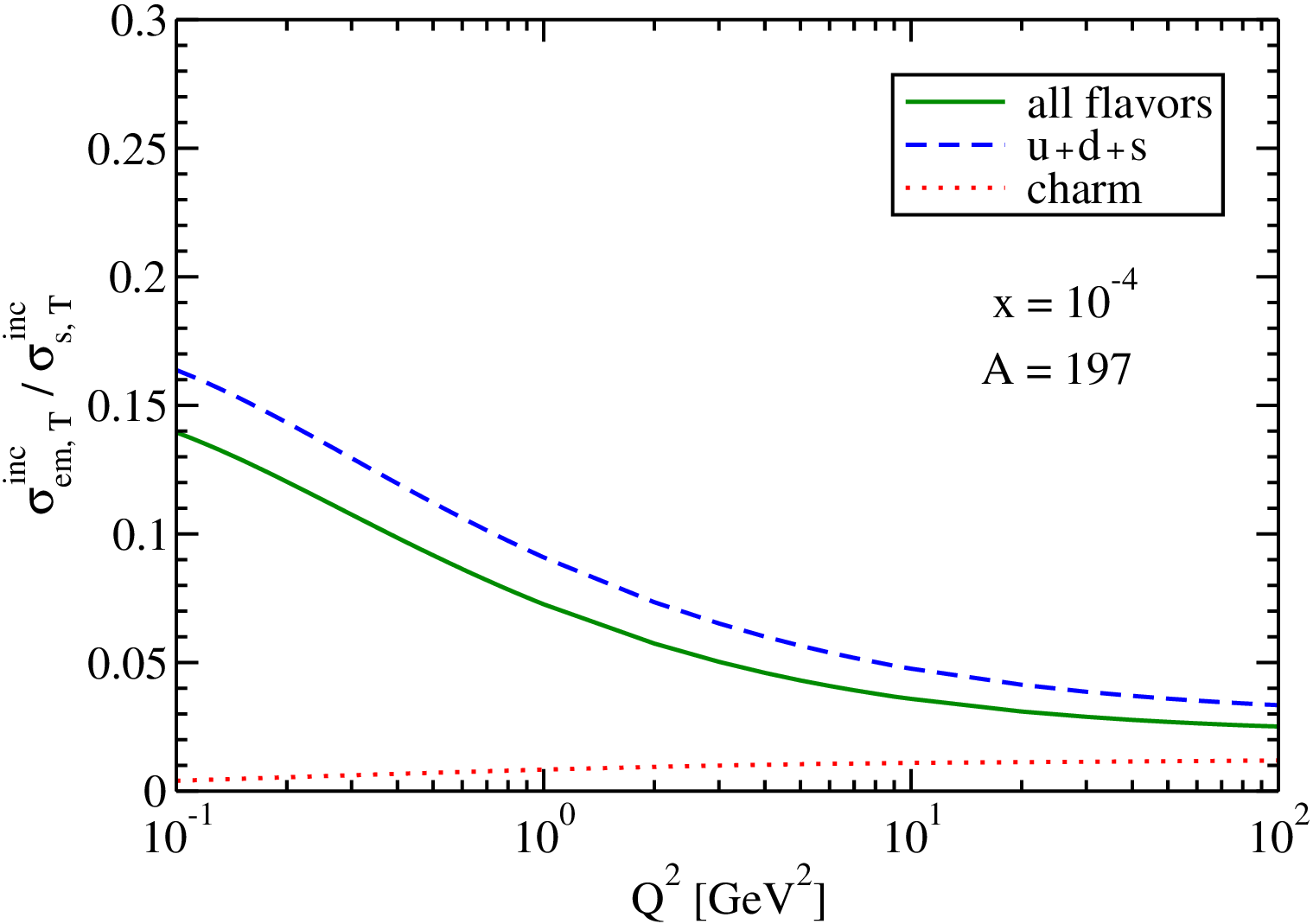}
  \includegraphics[scale=0.33]{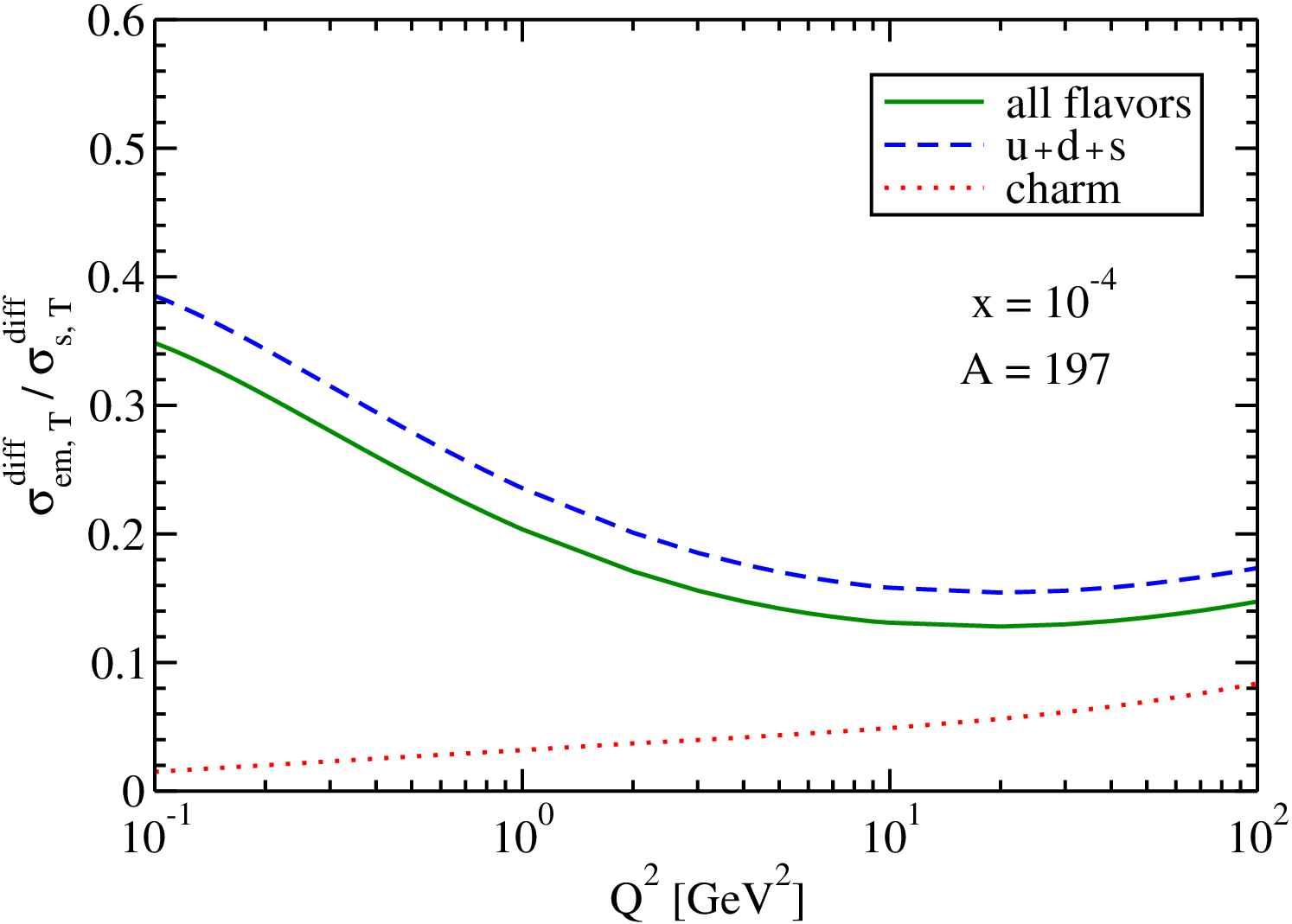}
\caption{ Upper panels: Dependence on $Q^2$ of the ratio between the electromagnetic and strong longitudinal  cross sections for the inclusive (left) 
and diffractive (right) interactions. Lower panels: Dependence on $Q^2$ of the ratio between the electromagnetic and strong transverse  cross sections 
for  inclusive (left) and diffractive (right) interactions.}
\label{fig4}
\end{figure}

\begin{figure}[t]
  \centering
  \includegraphics[scale=0.35]{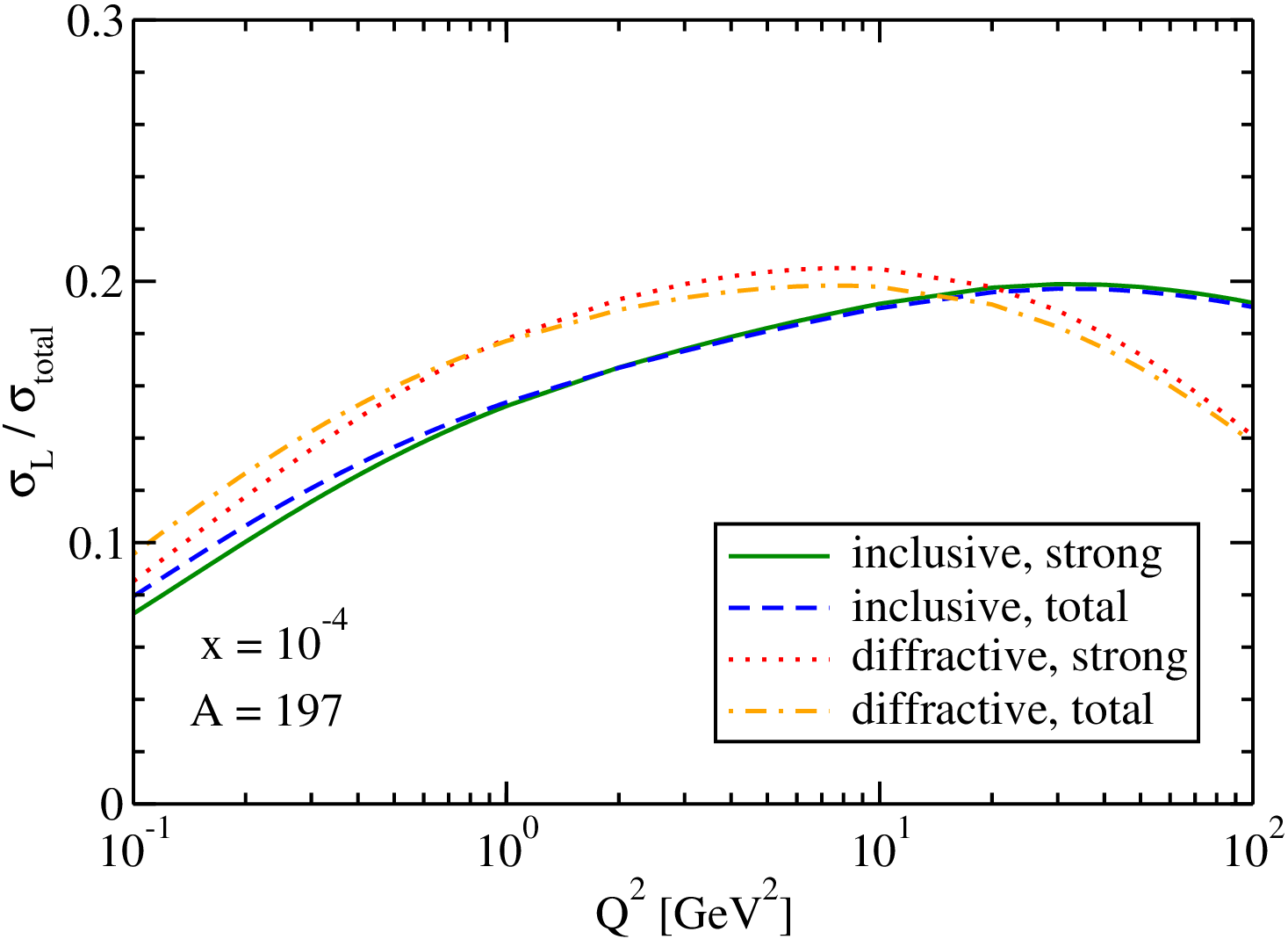}
  \includegraphics[scale=0.35]{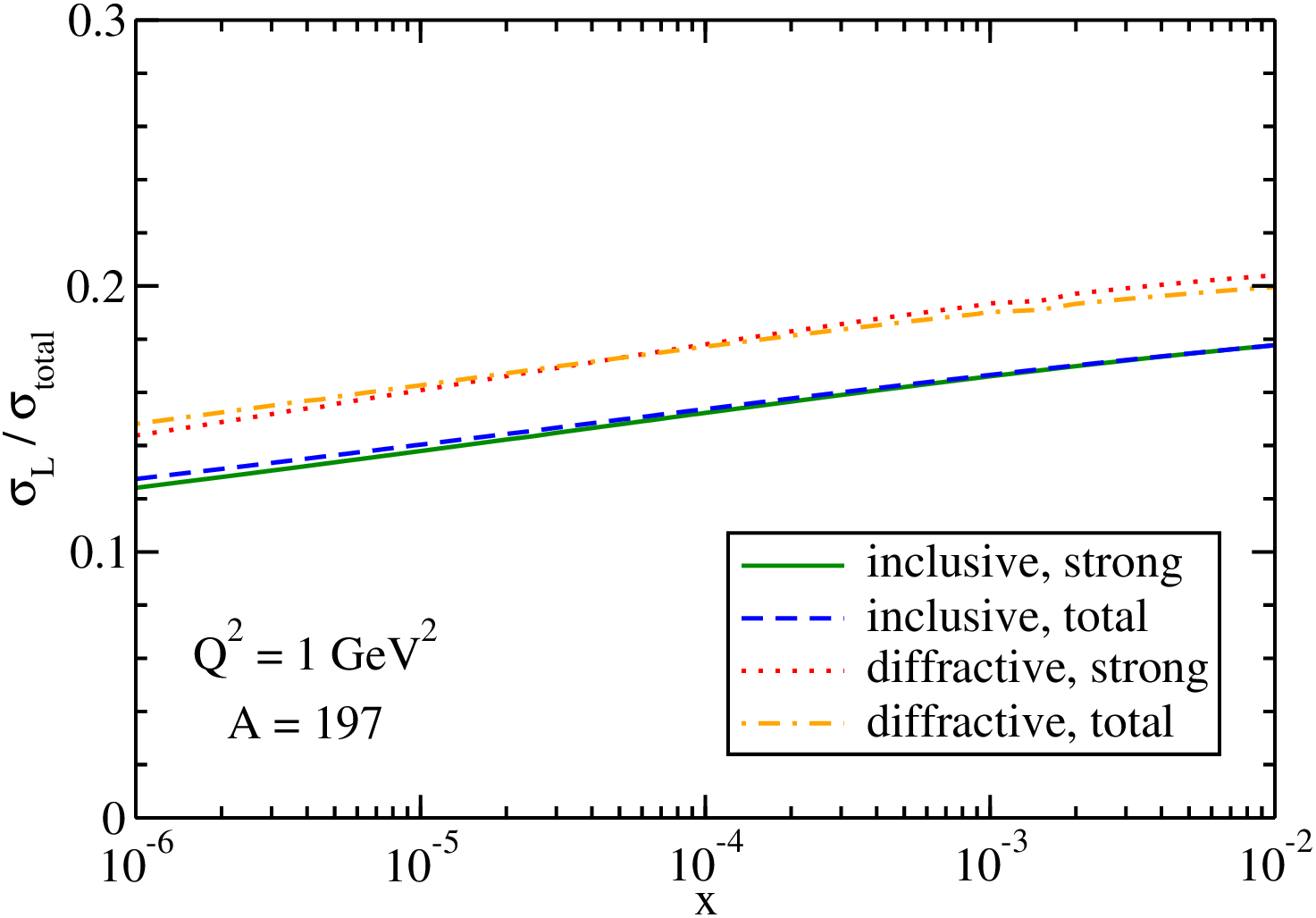}
 \caption{Dependence on $Q^2$ (left panel) and $x$ (right panel) for the ratio between the longitudinal and total cross sections for inclusive and diffractive 
interactions.}
\label{fig5}
\end{figure}

Let us now analyze more carefully the impact of the Coulomb corrections on the charm cross section, which is expected to be measured in the future 
electron - ion collider. In Fig. \ref{fig3} we present our predictions for the $Q^2$ (left panel) and $x$ (right panel) dependence  of the ratio
 between the charm and the total cross sections for inclusive and diffractive interactions. The predictions obtained disregarding the Coulomb corrections 
are denoted as {\it strong} in the figure. The charm contribution increases with $Q^2$ and at smaller values of $x$. Moreover, it is larger for inclusive 
processes. Our results indicate that the inclusion of the Coulomb corrections implies a mild decreasing of the ratios. In particular for inclusive 
processes, the $Q^2$ and $x$ dependence of the ratios are only slightly modified by Coulomb corrections in the perturbative $Q^2$ range that will probed 
in the future electron - ion collider. This result implies that the study of the charm production is a good probe of the high energy regime of the QCD dynamics.

Let us now consider the impact of the Coulomb corrections on the longitudinal and transverse cross sections in inclusive and diffractive interactions. 
In Fig. \ref{fig4} (upper panels) we present our results for the  ratio between the electromagnetic and strong longitudinal  cross sections. 
On the left (right) panel we show the inclusive (diffractive) cross sections.   In the inclusive case,  the ratio rapidly decreases with $Q^2$  and becomes 
smaller than 5 \% for $Q^2 \gtrsim 1.5$ GeV$^2$. In the diffractive case, the Coulomb correction is a factor 2 larger in the same kinematical range. As 
expected from our previous analysis, the charm contribution is  small in the $1 \le Q^2 \le 10$ GeV$^2$ 
range. The results presented in the lower panels of  Fig. \ref{fig4} are analogous to those of the upper panels but refer to the transverse cross sections. 
They indicate that the Coulomb corrections to both longitudinal and transverse cross sections are comparable.  Moreover, compared to the strong cross sections, 
they are small everywhere, except in the very low $Q^2$ ($Q^2 \simeq 0.1$ GeV$^2$) region, where dipoles of larger size dominate the cross sections.  In the figures 
we also see that for the  charm cross sections, the Coulomb corrections are very small. The fact that the longitudinal cross section is not sensitive to the 
Coulomb corrections is illustrated in a different way in Fig. \ref{fig5} where we present the $Q^2$ and $x$ dependence of the ratio $\sigma_L/\sigma_{tot}$ 
in inclusive and diffractive interactions. These results demonstrate that the inclusion of the Coulomb corrections has no impact on the ratio. The same is true 
for the transverse cross sections.  Therefore, the study of the longitudinal and transverse cross sections can also be useful to understand  the QCD dynamics 
at small - $x$.



A final comment is in order. In the analysis presented above we have assumed, following Refs. \cite{TuchinPRL,TuchinPRC}, that the modelling of the strong dipole -- nucleus scattering amplitude is given by Eq. (\ref{N_GBW}). However, as discussed in Refs. \cite{erike_ea2,vic_erike}, a more realistic model for  $\N_{s}(x,\rr,\rb)$ can be derived using the Glauber -- Mueller (GM) approach \cite{mueller}. In particular, in Ref. \cite{erike_ea2} two of the authors have demonstrated that the GM approach is able to describe  the current data for the nuclear structure function. In order to verify the dependence of our predictions on the model used to describe $\N_{s}$ we also have estimated the different observables discussed in this Section using the GM approach and  have obtained that the contribution of the Coulomb corrections  still are smaller than those presented here.

\section{Summary}
\label{conc}

In this paper we have studied in detail  the Coulomb contribution to some of the observables that will be measured in the future $eA$ collider. 
In particular, we study the impact of 
these corrections on the total, charm and longitudinal cross sections of inclusive and diffractive interactions. Our analysis is motivated by the fact 
that these observables are expected to probe the QCD dynamics and constrain the description of the non - linear effects. Our results indicate that the 
Coulomb corrections to the total cross sections are important at low - $Q^2$ and small values of $x$ and are larger for diffractive interactions. In 
particular, the ratio between the diffractive and total cross sections are enhanced by $\approx 10 \%$. In contrast, our results indicate the impact 
of the Coulomb corrections on the  transverse and longitudinal cross sections is small and it is negligible on the charm cross sections.  Therefore, these 
observables (especially the latter) can be considered  clean probes of the QCD dynamics.

\begin{acknowledgments}
The authors thank Kirill Tuchin for helpful comments. This work was  partially financed by the Brazilian funding
agencies CNPq, CAPES, FAPERGS, FAPESP (contract 12/50984-4) and  INCT-FNA (process number 
464898/2014-5).
\end{acknowledgments}


\end{document}